\documentclass[12pt]{iopart}
\usepackage{iopams}
\usepackage{graphicx}
\input amssym.def \input amssym

\begin{document}


\title[]{Three-qubit entangled embeddings of $CPT$ and Dirac groups within $E_8$ Weyl group}

\author{Michel Planat
}

\address{Institut FEMTO-ST, CNRS, 32 Avenue de
l'Observatoire,\\ F-25044 Besan\c con, France (planat@femto-st.fr)}

 

\begin{abstract}

In quantum information context, the groups generated by Pauli spin matrices, and Dirac gamma matrices, are known as the single qubit Pauli group $\mathcal{P}$, and two-qubit Pauli group $\mathcal{P}_2$, respectively. It has been found [M. Socolovsky, Int. J. Theor. Phys. 43, 1941 (2004)] that the $CPT$ group of the Dirac equation is isomorphic to $\mathcal{P}$. One introduces a two-qubit entangling orthogonal matrix $S$ basically related to the $CPT$ symmetry. With the aid of the two-qubit swap gate, the $S$ matrix allows the generation of the three-qubit real Clifford group and, with the aid of the Toffoli gate, the Weyl group $W(E_8)$ is generated (M. Planat, Preprint 0904.3691). In this paper, one derives three-qubit entangling groups $\tilde{\mathcal{P}}$ and $\tilde{\mathcal{P}}_2$, isomorphic to the $CPT$ group $\mathcal{P}$ and to the Dirac group $\mathcal{P}_2$, that are embedded into $W(E_8)$. One discovers a new class of pure three-qubit quantum states with no-vanishing concurrence and three-tangle that we name $CPT$ states. States of the $GHZ$ and $CPT$ families, and also chain-type states, encode the new representation of the Dirac group and its $CPT$ subgroup.

\end{abstract}

\pacs{03.67.Pp, 03.67.Pp, 02.20.-a, 03.65.Ud}


\section{Introduction}
In quantum field theory, a set of discrete transformations $T$ (a time reversal), $P$ (a space reversal, or parity) and $C$ (the charge conjugation) preserve the CPT symmetry. A $CPT$ violation would imply violation of Lorentz invariance. The mirror-image of the universe with objects reversed in time and space, and matter replaced by antimatter, would evolve like our universe. It is expected that $CPT$ invariance is a first principle that any physical theory should rely on.

Any discrete symmetry may be realized as a relation between two elements of a finite group. For the Dirac equation, the $CPT$ group $G$ is a subgroup of the group determined by $4 \times 4$ Dirac matrices. The relevant $CPT$ generators are (see \cite{Soco04}, eq. 37b).
\begin{equation}
P=i\gamma_0,~~  C=i\gamma_2\gamma_0~~\mbox{and}~~T=\gamma_3\gamma_1,
\label{eqCPT}
\end{equation} 
where the gamma matrices involved are \small $\gamma_0= \left(\begin{array}{cc} 1 &0 \\0& -1\end{array}\right)$, $\gamma_k= \left(\begin{array}{cc} 0 & \sigma_k \\- \sigma_k& 0\end{array}\right)$ \normalsize ($k=x$, $y$ and $z$), with \small $\sigma_x=\left(\begin{array}{cc} 0 & 1 \\1& 0\end{array}\right)$, $\sigma_y=\left(\begin{array}{cc} 0 & -i \\i         & 0\end{array}\right)$ and $\sigma_z=\left(\begin{array}{cc} 1 &0 \\0& -1\end{array}\right)$ \normalsize the Pauli spin matrices.

The $4\times 4$ matrices in (\ref{eqCPT}) generate a sixteen element group isomorphic to the small permutation group $\left\lfloor 16,13\right\rfloor$, a split group extension by $\mathbb{Z}_2$ of either of the three eight-element groups $\mathbb{Z}_2 \times \mathbb{Z}_4$, the dihedral group $D_4$ or the quaternion group $Q$. In other words, the $CPT$ group $G$ can be described by either of the semi-direct products
\begin{equation}
G\cong (\mathbb{Z}_2 \times \mathbb{Z}_4)\rtimes\mathbb{Z}_2\cong D_4\rtimes \mathbb{Z}_2\cong Q\rtimes \mathbb{Z}_2. 
\label{eqSD}
\end{equation} 
The $CPT$ group may also be seen as the central product $G\cong E_8^+ * \mathbb{Z}_4$, where $E_8^+ \equiv D_4$ is an extraspecial group of order $8$.
The group $G$ is isomorphic to the single qubit Pauli group $\mathcal{P}$, generated by Pauli spin matrices, i.e.
\begin{equation}
G\cong \mathcal{P} =\left\langle \sigma_x,\sigma_y,\sigma_z\right\rangle  \equiv G(4,2,2), 
\label{eq3}
\end{equation} 
where $G(4,2,2)$ is a imprimitive reflection group \\
(see http://en.wikipedia.org/wiki/Complex\_reflection\_group for the definition).

In this paper, one discovers another isomorphism to the $CPT$ group, that is constructed from a three-qubit entangled system. It was shown earlier \cite{Planat09} that the largest reflection group $W(E_8)$, of order $696~729~600$, can be enacted (i.e., represented) from a specific set of entangling matrices of the orthogonal group $SO(8)$. The construction follows from a building block $2 \times 2$ orthogonal matrix $S$, arising in Mermin's proof of the Kochen-Specker theorem. Basically, there are several entangled real three-qubit groups, that inflate to $W(E_8)$ under the action of the Toffoli gate $TOF$. Here one discovers that the smallest reflection subgroup that inflates to $W(E_8)$ under the action of $TOF$ is $\tilde{\mathcal{P}}$, the three-qubit representation of the Pauli group $\mathcal{P}$. 

Thus, the $CPT$ group is relevant in the context of $\mathcal{P}$, the Dirac equation, and the largest crystallographic group group $W(E_8)$. In Sec. 2, the $CPT$ matrix $S$ is introduced in relation to Mermin's study of quantum paradoxes. In Sec. 3, the three-qubit orthogonal representation of $\tilde{\mathcal{P}}$, that is isomorphic to the $CPT$ group is derived, and the corresponding entangled states, named $CPT$ states, are studied. In Sec. 4, the three-qubit representation of the Dirac group $\mathcal{P}_2$ is investigated.         

\section{Mermin's approach of quantum paradoxes and the $CPT$ matrix $S$}

The basic pieces of the proof of Kochen-Specker theorem in a four-dimensional space are two triples of (mutually commuting and real) two-qubit observables \cite{Planat09,Mermin93}
\begin{equation}
\left\{\sigma_x\otimes \sigma_x,\sigma_y\otimes \sigma_y,\sigma_z\otimes \sigma_z\right\}~\mbox{and}~\left\{\sigma_x\otimes \sigma_z,\sigma_z\otimes \sigma_x,\sigma_y\otimes \sigma_y\right\}.
\label{eq1}
\end{equation}
The joined eigenstates of the first triple of mutually commuting observables may be casted as the rows of the orthogonal matrix $R$ as below \cite{Planat09}
\begin{equation}
R=\frac{1}{\sqrt{2}}\left(\begin{array}{cccc} 1 & 0 & 0 & 1 \\0 & 1        & -1 & 0 \\ 0 & 1 & 1 & 0 \\-1 & 0 & 0 & 1\\ \end{array}\right),~~
\left(\begin{array}{ccc} + & + & - \\- &-        & -  \\ - & + &+  \\+ & - & + \\ \end{array}\right).
\end{equation}
Rows of the second matrix contain the sign of eigenvalues $\pm 1$, and each row corresponds to an entangled state, e.g. the state associated to the first row is $\frac{1}{\sqrt{2}}(\left|00\right\rangle+\left|11\right\rangle)$. The matrix $R$ occurs in the braiding approach of quantum computing \cite{Kauf04,PlanatJorrand08}. 

The joined eigenstates of the second triple of mutually commuting observables in (\ref{eq1}) may be similarly casted as the rows of the entangling orthogonal matrix
\begin{equation}
S=\frac{1}{2}\left(\begin{array}{cccc} 1 & -1 & 1 & 1 \\1 & 1        & -1 & 1 \\1 & -1 & -1 & -1 \\1 & 1 & 1 & -1\\ \end{array}\right),~~
\left(\begin{array}{ccc} + & - & - \\- &+        & -  \\ - & - & +  \\+ & + & + \\ \end{array}\right).
\end{equation}
The two matrices $R$ and $S$ capture, in a very compact form, the ingredients contained in the Mermin's proof of Kochen-Specker theorem.  The braid matrix $R$ satisfies the Yang-Baxter equation \cite{Kauf04}, but the $S$ matrix does not. Since $S$ is a building block of the new three-qubit realization of the $CPT$ group, one name it the $CPT$ matrix. Both matrices are related by a relation involving the Hadamard matrix $H$ as
$RS=H\otimes I$, with $H$ the Hadamard matrix.

Matrices $R$ and $S$ are used in the decomposition of the $n$-qubit Clifford group into {\it Clifford group dipoles} (see \cite{Planat09} for details about this terminology). The pair $(R,S)$ generates a group isomorphic to the reflection group $\mathcal{U}_{13}$, of order $96$, related to octahedral invariance \cite{Planat09}.  Only the $CPT$ matrix $S$ will play a role for the new representation of the $CPT$ group.

\section{Entanglement in the $CPT$ group}

As recalled at the previous section, the Pauli spin matrices generate the Pauli group $\mathcal{P}$, that is isomorphic to the $CPT$ group $G$. 
The two-qubit Pauli group $\mathcal{P}_2$ is generated by the two-fold tensor power of Pauli spin matrices. One gets $\mathcal{P}_2 \cong \left\lfloor 64,66\right]\cong E_{32}^+ * \mathbb{Z}_4$, where $E_{32}^+$ is an extraspecial group of order $32$. Another relevant isomorphism relating $\mathcal{P}$ to $\mathcal{P}_2$ is  $\mathcal{P}_2\cong\left\lfloor16,13\right\rfloor \rtimes \mathbb{Z}_2^2$, that singles out the normal subgroup isomorphic to $\mathcal{P}$.

Finally, $\mathcal{P}_2$ may also be seen as the group generated by the five $\gamma$ matrices occuring in the Dirac equation, that are  $\gamma_0$, $\gamma_k$ ($k=x$, $y$ and $z$) and  the chirality matrix $\gamma_5=\sigma_x \otimes 1$, with $1$ the $2 \times 2$ unity matrix. The first four $\gamma$ matrices generate a group isomorphic to the extraspecial group $E_{32}^-$ \footnote{The two extraspecial subgroups $E_{32}^+$ and $E_{32}^-$ of $\mathcal{P}_2$ are both normal in $\mathcal{P}_2$, and individually in the two-qubit Clifford group dipoles $\mathcal{C}_2^+$ and $\mathcal{C}_2^-$. See eq. (14) in \cite{Planat09}.}.

The first and second triple of observables in (\ref{eq1}) generate groups isomorphic to $\mathbb{Z}_2^3$ and $\mathbb{Z}_2^2$, respectively. The six observables in (\ref{eq1}) generate a group isomorphic to $\left\lfloor 16,11\right\rfloor \cong D_4 \times \mathbb{Z}_2$ (with $D_4$ the eight element dihedral group). Such a group was initially proposed as a tentative $CPT$ group of the Dirac equation in refs \cite{Soco04} and \cite{Varla04}.   

\subsection*{Derivation of the three-qubit representation of the $CPT$ group $\tilde{\mathcal{P}}$}

One way to arrive at the desired $3$-qubit representation $\tilde{\mathcal{P}}$ of the Pauli group $\mathcal{P}$ ( keeping in mind that the $CPT$ group $G$ is a $2$-qubit representation of $\mathcal{P}$) is to introduce the $3$-qubit {\it real} Clifford group $\mathcal{C}_3^+$ \cite{Nebe01}. In \cite{Planat09}, it is the {\it real} dipole of the $3$-qubit complex Clifford group $\mathcal{C}_3$ and is represented as
\begin{equation}
\mathcal{C}_3^+=\left\langle 1 \otimes S, S\otimes 1, 1 \otimes T, T\otimes 1\right\rangle,
\end{equation}
with four generators, comprising the $CPT$ matrix $S$ and the swap matrix $T$ in their factors.
Group $\mathcal{C}_3^+$, of order $2~580~480$, may be inflated to a representation of the largest reflection group $W(E_8)$, by adjoining to it the Toffoli gate generator $TOF=C^2NOT$ [Recall that $CNOT$  gate flips the second qubit (the target qubit) if and only if the first qubit (the control qubit) is 1.]    The group $\mathcal{C}_3^+$ is isomorphic to a maximal subgroup of $W'(E_8)$.  
As observed for the first time in \cite{Planat09}, many reflection subgroups such as the Weyl group of $E_6$ and $E_7$, as well as groups $PSL(2,7)$ and $SL(2,5)$, inflate to $W(E_8)$ by adding the Toffoli gate generator.

Then, comes the question to determine the {\it kernel} of these inflations, i.e. the smallest subgroup inflating to $W(E_8)$ under the action of the Toffoli gate \footnote {Fredkin gate also does the job when substituted to the Toffoli gate.}.

The three-qubit $CPT$ group so defined reads $\tilde{\mathcal{P}}=\left\langle K,i,j\right\rangle$, where $\left\langle i,j\right\rangle \cong Q$ and 
$\left\langle K,i\right\rangle \cong D_4$, with generators  
\small
\begin{eqnarray}
&i=\frac{1}{2}\left(\begin{array}{cccccccc} 0&-1&1&0 &0&0&-1&-1 \\1&0&0&-1 &1&1&0&0\\-1&0&0&-1 &-1&1&0&0\\0&1&1&0 &0&0&1&-1\\
0&-1&1&0 &0&0&1&1 \\0&-1&-1&0 &0&0&1&-1\\1&0&0&-1 &-1&-1&0&0\\1&0&0&1 &-1&1&0&0\\ \end{array}\right), \nonumber \\
&j=\frac{1}{2}\left(\begin{array}{cccccccc} 0&1&1&1 &1&0&0&0 \\-1&0&0&0 &0&1&1&-1\\-1&0&0&0 &0&-1&-1&-1\\-1&0&0&0 &0&-1&1&1\\
-1&0&0&0 &0&1&-1&1 \\0&-1&1&1 &-1&0&0&0\\0&-1&1&-1 &1&0&0&0\\0&1&1&-1 &-1&0&0&0\\ \end{array}\right),\nonumber \\
&K=\frac{1}{2}\left(\begin{array}{cccccccc} 1&0&0&0 &0&1&1&1 \\0&-1&1&1 &-1&0&0&0\\0&1&1&1 &1&0&0&0\\0&1&1&-1 &-1&0&0&0\\
0&-1&1&-1 &1&0&0&0 \\1&0&0&0 &0&-1&-1&1\\1&0&0&0 &0&-1&1&-1\\1&0&0&0 &0&1&-1&-1\\ \end{array}\right). 
\label{CPTgroup}
\end{eqnarray}
\normalsize
\subsection*{Measures of entanglement}

The resources needed to create a given entangled state may be quantified, and one can define invariants for discriminating the type of entanglement.

For a pair of quantum systems $A$ and $B$ in a pure state of density matrix $\left|\psi \right\rangle \left\langle \psi \right|$, the {\it entanglement of formation} is defined as the entropy of either of the two subsystems $A$ and $B$
\begin{equation}
E(\psi)=-\mbox{tr}(\rho_A \log_2 \rho_A)=-\mbox{tr}(\rho_B \log_2 \rho_B),
\end{equation}
where $\rho_A$ and $\rho_B$ are partial traces of $\rho$ over subsystems $B$ and $A$, respectively.
The measure is made explicit by defining the spin-flipped density matrix \cite{Wootters00} 
\begin{equation}
\tilde{\rho}=(\sigma_y \otimes \sigma_y)\rho^{\ast}(\sigma_y \otimes \sigma_y), 
\end{equation}
and the concurrence $C(\psi)=|\langle \psi |\tilde{\psi}\rangle|$ between the original and flipped state $\tilde{\psi}=\sigma_y \left|\psi^{\ast}\right\rangle$. As both $\rho$ and $\tilde{\rho}$ are positive operators, the product $\rho\tilde{\rho}$ also has only real and non-negative eigenvalues $\lambda_i$ (ordered in decreasing order) and the concurrence reads 
\begin{equation}
C(\rho)=\mbox{max}\left\{0,\sqrt{\lambda_1}-\sqrt{\lambda_2}-\sqrt{\lambda_3}-\sqrt{\lambda_4}\right\}. 
\end{equation}

For a two-qubit state  $\left| \psi\ \right\rangle=\alpha \left|00\ \right\rangle + \beta\left|01\ \right\rangle+ \gamma\left|10\ \right\rangle+ \delta\left|11\ \right\rangle$, the concurrence is $C=2\left|\alpha\delta-\beta\gamma\right|$, and thus satisfies the relation $0\le C \le 1$, with $C=0$ for a separable state and $C=1$ for a maximally entangled state.

The entanglement of a triple of quantum systems $A$, $B$ and $C$ in a pure state may be conveniently described by tracing out over partial subsystems $AB$, $BC$, and $AC$. In this generalized context, one introduces the {\it tangle} $\tau=C^2$. Tangles attached to the bipartite subsystems above satisfy the inequality

\begin{equation}
\tau_{AB}+\tau_{AC}\le 4 \mbox{det}\rho_A \equiv \tau_{A(BC)}. 
\end{equation}
The right hand side is interpreted as the amount of entanglement shared by the single qubit $A$ with the pair $BC$, in comparison with the amounts of entanglement shared with qubits $B$ and $C$ taken individually. It is remarkable that, for any value of the tangles satisfying this inequality, one can find a quantum state consistent with those values \cite{Wootters00}.

It has been shown that an arbitrary three-qubit state $\left|\psi\right\rangle$ can be entangled in essentially two inequivalent ways, belonging to the GHZ-class: $\left|\mbox{GHZ}\right\rangle=\frac{1}{\sqrt{2}}(\left|000\right\rangle+\left|111\right\rangle)$ or to the W-class: $\left|\mbox{W}\right\rangle=\frac{1}{\sqrt{3}}(\left|001\right\rangle)+\left|010\right\rangle+\left|100\right\rangle)$, according whether $\psi$ can be converted to the state $\left|\mbox{GHZ}\right\rangle$ or to the state $\left|\mbox{W}\right\rangle$, by stochastic local operations and classical communication (SLOCC) \cite{Dur00}. The relevant class is determined by computing the bipartite tangles of the reduced subsystems. If they vanish, then the subsystems are separable and
$\left|\psi\right\rangle$ belongs to the GHZ-class, meaning that all the entanglement is destroyed by tracing over one subsystem. If none of the bipartite tangles vanish, then $\left|\psi\right\rangle$ belongs to the W-class, meaning that it maximally retains bipartite entanglement after tracing over one subsystem.

Further discrimination of the entanglement type of a general $3$-qubit state
\begin{equation}
\left|\psi\right\rangle=\sum_{a,b,c=0,1}\psi_{abc}\left|abc\right\rangle, 
\end{equation}
can be obtained by calculating the SLOCC invariant three-tangle \cite{Wootters00}
\begin{eqnarray}
&\tau^{(3)}=4\left|d_1-2d_2+4d_3\right|, \nonumber \\
&d_1=\psi_{000}^2\psi_{111}^2+\psi_{001}^2\psi_{110}^2+\psi_{010}^2\psi_{101}^2+\psi_{100}^2\psi_{011}^2,\nonumber \\
&d_2=\psi_{000}\psi_{111}(\psi_{011}\psi_{100}+\psi_{101}\psi_{010}+\psi_{110}\psi_{001})\nonumber \\
&+\psi_{011}\psi_{100}(\psi_{101}\psi_{010}+\psi_{110}\psi_{001})+\psi_{101}\psi_{010}\psi_{110}\psi_{001},\nonumber \\
&d_3=\psi_{000}\psi_{110}\psi_{101}\psi_{011}+\psi_{111}\psi_{001}\psi_{010}\psi_{100}. 
\end{eqnarray}
For the GHZ state the $3$-tangle becomes maximal: $\tau^{(3)}=1$ and it vanishes for any factorized state. It also vanishes for states of the $W$-class.
The $3$-tangle may be interpreted as the {\it residual tangle}
\begin{equation}
\tau^{(3)}=\tau_{A(BC)}-(\tau_{AB}+\tau_{AC}), 
\end{equation}
i.e., the amount of entanglement between subsystems $A$ and $BC$ that cannot be accounted for by the entanglements of $A$ with $B$ and $C$ separately. It is of course independent on which qubit one takes as the reference of the construction. The GHZ state is a true tripartite entangled state so that no amount of entanglement is in the bipartite subsystems, as a result the residual entanglement is maximal. In contrast, for the states of the W-class the entanglement is of a pure bipartite type and $\tau^{(3)}=0$. Mixtures of GHZ and W states are studied in \cite{Lohmayer06}, where it is shown that while the amounts of inequivalent entanglement types strictly add up for pure states, the {\it monogamy} is in general lifted for mixed states because the entanglement can arise from different types of locally inequivalent quantum correlations.

Knowing the three-tangle $\tau^{(3)}$ and the two tangles $\tau_{AB}$ and $\tau_{AC}$ of subsystems $AB$ and $AC$, the linear entropy (one-tangle) $\tau_{A(BC)}$ may also be calculated \cite{Wootters00,Lohmayer06}. It is a measure of the full amount of entanglement in the system and for a mixed three-qubit state it may take a non-zero value even if no two- and three-partite entanglement is present. 

Let us investigate the type and amount of entanglement in the three-qubit pure states  $\left|CPT\right\rangle$ arising from the $CPT$ group $\tilde{\mathcal{P}}$. 
One singles out the state arising from the first row of the generator $K$. The same measures are obtained for states arising from the quaternion generators  $i$ or $j$. For the state 

\begin{equation}
\left|CPT\right\rangle=\frac{1}{2}(\left|000\right\rangle+\left|101\right\rangle+\left|110\right\rangle+\left|111\right\rangle),
\label{CPT}
\end{equation}
the three-tangle is $\tau^{(3)}=\frac{1}{4}.$ 

The density matrices of the bipartite subsystems are 
\scriptsize
$$\rho_{BC}=\frac{1}{4}\left(\begin{array}{cccc} 1 & 0 & 0 & 0 \\0 & 1 & 1 & 1 \\ 0 & 1 & 1 & 1 \\0 & 1 & 1 & 1\\ \end{array}\right),~
\rho_{AB}=\frac{1}{4}\left(\begin{array}{cccc} 1 & 0 & 0 & 1 \\0 &0 & 0 & 0 \\ 0 & 0 & 1 & 1 \\1 & 0 & 1 & 2\\ \end{array}\right),~
\rho_{AC}=\frac{1}{4}\left(\begin{array}{cccc} 1 & 0 & 0 & 1 \\0 & 0 & 0 & 0 \\ 0 & 0 & 1 & 1 \\1 & 0 & 1 & 2\\ \end{array}\right).~$$
%
\normalsize

The set of eigenvalues $\left\{\frac{1}{16}(3+2\sqrt{2}), \frac{1}{16}(3-2\sqrt{2}),0,0\right\}$ is uniform over the subsystems. All $CPT$ states exhibit the same entanglement measures $\tau^{(3)}=\frac{1}{4}$, $\tau_{AB}=\tau_{AC}=\tau_{BC}=\frac{1}{4}$.  Thus, the entanglement measure for two parties equals the entanglement measure for three parties. The linear entropy is
\begin{equation}
\tau_{A(BC)}= \frac{1}{4}+ 2\frac{1}{4}=\frac{3}{4}.
\end{equation}

It is tempting to compare a $CPT$ state as in (\ref{CPT}) state to the unique mixed state 
\begin{equation}
\left|Z\right\rangle=\sqrt{p}\left|GHZ\right\rangle-e^{-i \phi}\sqrt{1-p}\left|W\right\rangle
\end{equation} 
with the same three-tangle $\tau^{(3)}=\frac{1}{4}$. According to Fig. 3 in \cite{Lohmayer06}, one gets for such a mixed state 
$p \approx 0.70$, the sum of two concurrences $\tau_{AB}+\tau_{AC}\approx 0$ and $\tau_{A(BC)}\approx 0.85$. Clearly the CPT state and the $\left|Z\right\rangle$ state with the same three-tangle are completely different objects. 

\section{Entanglement in the Dirac group}

\subsection*{The three-qubit representation of the group design $SL(2,5)$}

Among the various groups, that can be inflated to $W(E_8)$ under the action of the Toffoli gate, the selection of $H_{120}=\tilde{SL}(2,5)$ is justified in many respects. First, a matrix representation of $SL(2,5)$ is a unitary design, i.e. a set of unitary matrices that simulates the entire unitary group \cite{Gross07}. A two-dimensional complex representation of $SL(2,5)$ is given in \cite{Planat09}, eq. (15). The three-dimensional orthogonal representation is as below. The group design $SL(2,5)$ is the smallest known $2$-dimensional $5$-design.

Second, recall that the Poincar\'e dodecahedral space $\mathcal{D}$ is a tentative model of the far universe, that describes well the fluctuations of the cosmic microwave background \cite{Weeks06}. The fundamental group of $\mathcal{D}$ is the binary icosahedral group, isomorphic to $SL(2,5)$.    

One gets 
\begin{equation}
\langle x,y, TOF\rangle\cong W(E_8)~\mbox{with}~\mathcal{C}_3^+=\langle x,y,1 \otimes CZ \rangle~\mbox{and}~ \langle x,y\rangle \cong SL(2,5),
\label{SL25}
\end{equation}
where $CZ=\mbox{diag}(1,1,1,-1)$ and the generators are
\small
\begin{eqnarray}
&x=\frac{1}{2}\left(\begin{array}{cccccccc} 1&-1&0&0 &0&0&1&1 \\1&1&0&0 &0&0&-1&1\\0&0&1&1 &-1&1&0&0\\0&0&-1&1 &-1&-1&0&0\\
0&0&1&1 &1&-1&0&0 \\0&0&-1&1 &1&1&0&0\\-1&1&0&0 &0&0&1&1\\-1&-1&0&0 &0&0&-1&1\\ \end{array}\right),\nonumber \\
&y=\frac{1}{2}\left(\begin{array}{cccccccc} 1&0&-1&0 &0&-1&0&-1 \\0&1&0&1 &1&0&-1&0\\1&0&1&0 &0&-1&0&1\\0&-1&0&1 &-1&0&-1&0\\
0&-1&0&1 &1&0&0&1 \\1&0&1&0 &0&1&0&-1\\0&1&0&1 &-1&0&0&1\\1&0&-1&0 &1&0&0&1\\ \end{array}\right)\nonumber .\\
\end{eqnarray}

\normalsize

It is straightforward to calculate the invariants attached to  states of the type
\begin{eqnarray}
\left|\psi\right\rangle=\frac{1}{2}(\left|000\right\rangle-\left|001\right\rangle+\left|110\right\rangle+\left|111\right\rangle), \\ \nonumber
\mbox {or}~~\left|\psi\right\rangle=\frac{1}{2}(\left|000\right\rangle-\left|010\right\rangle-\left|101\right\rangle-\left|111\right\rangle),
\label{SL(2,5)}
\end{eqnarray}
that correspond to the first rows of $x$ and $y$, respectively. The three-tangle of the states are $\tau^{(3)}=1$ and the two-partite density matrices uniformly possess the set of square eigenvalues $\left\{\frac{1}{4},\frac{1}{4},0,0\right\}$ corresponding to vanishing concurrence.  Thus, the entanglement arising from the generators of $H_{120}$ is of the $GHZ$ type. 

The group $H_{120}$ expands in size by adding to it one of the generators of $\tilde{\mathcal{P}}$. Adding the quaternionic generator $i$ to $H_{120}$, one gets a group isomorphic to $E_{32}^-.S_5$ that is, up to a factor of $2$, the $3$-qubit representation of the $2$-qubit dipole $\mathcal{C}_2^-$, of order $3840$ (see eq. (14) in \cite{Planat09}). Adding the quaternionic generator $j$ to $H_{120}$, one recovers the $3$-qubit representation of the group $\mathbb{Z}_2.W'(E_6)$, and adding the generator $K$ to $H_{120}$ one obtains the group $\tilde{F}_7 \cong \mathcal{Z}_2.W'(E_7)$\footnote{The group $W'(E_8)$ contains three maximal subgroups of order $2~903~040$. One of them is isomorphic to $W(E_7)$ and the remaining two are isomorphic to $\tilde{F}_7$. This is the largest size for a maximal subgroup of $W'(E_8)$. The second largest size for a  maximal subgroup of $W'(E_8)$ is $2~580~480$. One of the three maximal subgroups of this size is isomorphic to the real Clifford group $\mathcal{C}_3^+$, as already mentioned.}.

\subsection*{Entanglement in the extraspecial group $E_{32}^-$}
\label{extra}

In (\ref{CPTgroup}), the quaternion group was generated with the two $CPT$ generators $i$ and $j$. It can also be obtained by using two non-$CPT$ generators 
$W$ and $Z$ , i.e. $\left\langle W,Z\right\rangle \cong Q$ with
\small
$$W=\frac{1}{2}\left(\begin{array}{cccccccc} 0&0&0&-1 &0&1&-1&1 \\0&0&0&-1 &0&-1&1&1\\0&0&0&-1 &0&1&1&-1\\1&1&1&0 &1&0&0&0\\
0&0&0&-1 &0&-1&-1&-1 \\-1&1&-1&0 &1&0&0&0\\1&-1&-1&0 &1&0&0&0\\-1&-1&1&0 &1&0&0&0\\ \end{array}\right),$$

$$Z=\frac{1}{2}\left(\begin{array}{cccccccc} 0&-1&0&-1 &1&-1&0&0 \\1&0&1&0 &0&0&1&-1\\0&-1&0&1 &1&1&0&0\\1&0&-1&0 &0&0&-1&-1\\
-1&0&-1&0 &0&0&1&-1 \\1&0&-1&0 &0&0&1&1\\0&-1&0&1 &-1&-1&0&0\\0&1&0&1 &1&-1&0&0\\ \end{array}\right).$$

\normalsize	

The states arising from the generator $W$ are maximally entangled and of the type $W$, i.e. $\tau^{(3)}=0$, the concurrences of the subsystems equal $\frac{1}{2}$ and the linear entropy $\tau_{A(BC)}$ equals $1$. The states arising from the generator $Z$ are such that $\tau^{(3)}=\frac{1}{4}$, $\tau_{AB}=\tau_{AC}=\frac{1}{4}$ and $\tau_{BC}=0$. They are of the chain-type $B-A-C$.

of the $GHZ$ type. 

Then, using the following $CPT$ matrix

$$c=\frac{1}{2}\left(\begin{array}{cccccccc} 1&0&-1&-1 &0&1&0&0 \\0&-1&0&0 &-1&0&-1&-1\\-1&0&1&-1 &0&1&0&0\\-1&0&-1&1 &0&1&0&0\\
0&-1&0&0 &-1&0&1&1 \\1&0&1&1 &0&1&0&0\\0&-1&0&0 &1&0&-1&1\\0&-1&0&0 &1&0&1&-1\\ \end{array}\right),$$

one gets a new realization of the $CPT$ group
\begin{equation}
\left\langle W,Z,c\right\rangle \cong \left\lfloor 16,13\right\rfloor
\label{newCPT}
\end{equation}

One also introduces another matrix $Z'$, encoding states of the chain-type $A-B-C$  

$$Z'=\frac{1}{2}\left(\begin{array}{cccccccc} 0&0&-1&0 &-1&-1&-1&0 \\0&0&1&0 &1&-1&-1&0\\1&-1&0&-1 &0&0&0&-1\\0&0&1&0 &-1&1&-1&0\\
1&-1&0&1 &0&0&0&-1 \\1&1&0&-1 &0&0&0&-1\\1&1&0&1 &0&0&0&1\\0&0&-1&0 &1&1&-1&0\\ \end{array}\right).$$

Using generators $W$, $Z$ and $Z'$, one obtains a modified $CPT$ group as  
\begin{equation}
\left\langle W,Z,Z'\right\rangle \cong \left\lfloor 16,12\right\rfloor=Q \times \mathbb{Z}_2.
\label{mCPT}
\end{equation}

In \cite{Soco04}, it is shown that the $CPT$ group of the {\it Dirac field} (not of the Dirac equation), which acts on the Hilbert space of the field theory, is isomorphic to the group $Q \times \mathbb{Z}_2$. The $3$-qubit representations $(\ref{newCPT})$ and $(\ref{mCPT})$ immediately leads to the group encompassing the $CPT$-group $\left\lfloor 16,13\right\rfloor$ of the Dirac equation and the $CPT$ group $\left\lfloor 16,12\right\rfloor$ of the Dirac field as   
\begin{equation}
\left\langle W,Z,c,Z'\right\rangle \cong \left\lfloor 32,50\right\rfloor\equiv E_{32}^-.
\label{fullCPT}
\end{equation}
The extraspecial group $E_{32}^-$ is isomorphic to the unique normal subgroup of order $32$ of the Clifford group dipole $\mathcal{C}_2^-$, and also corresponds to the group generated by the first four $\gamma$ matrices. 

\subsection*{Entanglement in the Dirac group}

To arrive at the expected $3$-qubit representation $\tilde{\mathcal{P}}_2$ of the Dirac group $\mathcal{P}_2$, one adds to the representation $\tilde{SL}(2,5)$, given in (\ref{SL25}), the generator $K$, given in (\ref{CPTgroup}). This generates the group $\tilde{F}_7$. One the maximal subgroups of $\tilde{F}_7$, of order $46080$, is isomorphic to the non-split product $M=\mathcal{P}_2. S_6$, of order $46080$ (with $S_6$ the symmetric group on six letters), and the corresponding normal subgroup is represented as
\begin{equation}
\tilde{\mathcal{P}}_2=\left\langle g_1,g_2,c_1,c_2,u\right\rangle,
\end{equation}
with two $GHZ$-type generators $g_1$ ang $g_2$
\small
\begin{eqnarray}
&g_1= \frac{1}{2}\left(\begin{array}{cc} R_1 &R_2 \\R_2& R_1\end{array}\right),~~g_2= \frac{1}{2}\left(\begin{array}{cc} R_1 &-R_2 \\-R_2& R_1\end{array}\right),~\mbox{with}  \nonumber \\
&R_1=\left(\begin{array}{cccc} -1 & 0 & 0 & 1 \\0 & 1        & -1 & 0 \\ 0 & -1 & 1 & 0 \\1 & 0 & 0 & -1\\ \end{array}\right)~~\mbox{and}~~R_2=\left(\begin{array}{cccc} 1 & 0 & 0 & 1 \\0 & -1        & -1 & 0 \\ 0 & -1 & -1 & 0 \\1 & 0 & 0 & 1\\ \end{array}\right),
\end{eqnarray}
\normalsize
two generators $c_1$ and $c_2$ of chain-type $B-A-C$ 
\footnotesize
\begin{eqnarray}
&c_1=\frac{1}{2}\left(\begin{array}{cccccccc} 0&1&0&-1 &-1&-1&0&0 \\-1&0&1&0 &0&0&1&1\\0&-1&0&1 &-1&-1&0&0\\1&0&-1&0 &0&0&1&1\\
1&0&1&0 &0&0&-1&1 \\1&0&1&0 &0&0&1&-1\\0&-1&0&-1 &1&-1&0&0\\0&-1&0&-1 &-1&1&0&0\\ \end{array}\right),\nonumber \\
&
c_2=\frac{1}{2}\left(\begin{array}{cccccccc} 0&1&0&1 &0&0&-1&1 \\-1&0&-1&0 &1&-1&0&0\\0&1&0&1 &0&0&1&-1\\-1&0&-1&0 &-1&1&0&0\\
0&-1&0&1 &0&0&1&1 \\0&1&0&-1 &0&0&1&1\\1&0&-1&0 &-1&-1&0&0\\-1&0&1&0 &-1&-1&0&0\\ \end{array}\right),
\end{eqnarray}
\normalsize
and the unentangled generator
\begin{eqnarray}
&u= \left(\begin{array}{cc} -U_1 &0 \\0& -U_2\end{array}\right),\nonumber \\
&\mbox{with}~~U_1=\left(\begin{array}{cccc} 0 & 1 & 0 & 0 \\1 & 0 & 0 & 0 \\ 0 & 0 &0 & 1 \\0 & 0 & 1 & 0\\ \end{array}\right)~\mbox{and}~U_2=\left(\begin{array}{cccc} 0 & 0 & 1 & 0 \\0 & 0 & 0 & 1 \\ 1 & 0 &0 & 0 \\0 & 1 & 0 & 0\\ \end{array}\right).
\end{eqnarray}

Let us list a few subgroups of $\tilde{\mathcal{P}}_2$, that helps to clarify its physical structure. First, the pair of $GHZ$-type generators generates the Klein four group: $\left\langle g_1, g_2 \right\rangle \cong \mathbb{Z}_2^2$ , and the pair of chain-type generators generates the quaternion group:   $\left\langle c_1, c_2  \right\rangle \cong Q$. 

Second, by removing the $GHZ$-type generators $g_1$, one recovers a representation of the normal (extraspecial) group $E_{32}^-$. By removing either of the chain-type generators $c_1$, $c_2$, or the $GHZ$-type generator $g_2$, or the unentangled generator $u$, one arrives at a representation of the normal (extraspecial) subgroup $E_{32}^+$.

As a result, the $CPT$ group of the Dirac equation is obtained by removing $u$ from $E_{32}^-$
\begin{equation}
\left\langle g_1,c_1,c_2\right\rangle \cong \tilde{\mathcal{P}} \cong \left\lfloor 16,13\right\rfloor,
\end{equation}
and by removing $u$ from $E_{32}^+$ one gets  a group isomorphic to the $CPT$ group of the Dirac field
\begin{equation}
\left\langle g_2,c_1,c_2\right\rangle \cong \left\lfloor 16,12\right\rfloor,
\end{equation}
or {\it the false} $CPT$ group.

\begin{equation}
\left\langle g_1,g_2,c_1\right\rangle \cong\left\langle g_1,g_2,c_2\right\rangle \cong \left\lfloor 16,11\right\rfloor.
\end{equation}
In \cite{Soco04}, the group $\left\lfloor 16,11\right\rfloor \equiv D_4 \times \mathbb{Z}_2$ is denoted $G_{\theta}^{1}$ and the $CPT$ group of the Dirac equation $\left\lfloor 16,13\right\rfloor \cong D_4 \rtimes \mathbb{Z}_2$ is denoted $G_{\theta}^{2}$. Both groups satisfy the requirement of $CPT$ invariance within Dirac equation. But, the consistency between the one particle Dirac theory and the quantum field theory selects the second solution. The group associated to the Dirac field is denoted $G_{\theta}\cong \left\lfloor 16,12\right\rfloor$ in \cite{Soco04}. Thus, the normal series $\tilde{Q} \triangleleft\tilde{\mathcal{P}} \triangleleft \tilde{E}_{32}^-  \triangleleft \tilde{\mathcal{P}}_2 \triangleleft \tilde{M} \subset \tilde{F}_7)$ helps to clarify the relevance of various groups (the tilde symbol means that we are dealing with the $3$-qubit representation). 

\section{Discussion}

Three-qubit entanglement, and its relationship to the largest crystallographic group $W(E_8)$, uncovered in this paper, is expected to play a role in two separate contexts: quantum computing and unifying approaches of physics. The single qubit Pauli group $\mathcal{P}$, the $CPT$ subgroup $G$ of the Dirac group and the kernel of entanglement $\tilde{\mathcal{P}}$ in the new three-qubit representation of $W(E_8)$ were found to be isomorphic. That may be a coincidence or the symptom of a more intricate physical theory, such as string theory \cite{Witten87}. Such a theory would encompass Dirac equation and be an alternative to quantum field theory. To conclude, quantum entanglement in $e^+e^-$ collisions was recently observed in relation to a possible $CPT$ violation  \cite{Go07}. 

\section*{Acknowledgements} The author is indepted to Miguel Socolovsky for his inspiring paper and for his careful reading of the manuscript. He also  acknowledges Maurice Kibler for his comments and his invitation to present this work at the Institut of Physique Nucl\'eaire in Lyon.

\end{document}